\documentclass[aps,prl,twocolumn,showpacs]{revtex4}

\usepackage{amsmath,amssymb,amsfonts,amsthm,bbm} 
\usepackage{color}
\usepackage{graphicx}
\usepackage[hang,nooneline]{subfigure}

\begin{document}

\title{Analytic solutions of the geodesic equation in axially symmetric space--times
}

\author{E. Hackmann$^1$, V. Kagramanova$^2$, J. Kunz$^2$ and C. L\"ammerzahl$^1$}

\affiliation{$^1$ ZARM, Universit\"at Bremen, Am Fallturm, 28359 Bremen, Germany}
\affiliation{$^2$ Institut f\"ur Physik, Universit\"at Oldenburg, 26111 Oldenburg, Germany}

\date\today

\begin{abstract}

The complete sets of analytic solutions of the geodesic equation in Taub--NUT--(anti-)de Sitter, Kerr--(anti-)de Sitter and also in general Pleba\'nski--Demia\'nski space--times without acceleration are presented. The solutions are given in terms of the Kleinian sigma functions. 

\end{abstract}


\maketitle

\section{Introduction and Motivation}

Supernovae and other data indicate the existence of Dark Energy which is appropriately described by a cosmological constant. Galactic rotation curves, gravitational lensing and structure formation indicate the existence of Dark Matter which in some approaches is accounted for by assuming a modified gravity theory. In any case, properties of the gravitational field, of the underlying geometry, can only be explored through the motion of test objects like point particles and light rays. Therefore analytic solutions of  the geodesic equation giving the orbits of test objects in standard and generalized gravitational models is an important issue. 

Analytic solutions of the geodesic equation are also of use for black hole physics and the creation of gravitational waves: they provide an analytic determination of the shape of black holes with a link to the black hole parameters \cite{HiokoMaeda09}, also in generalized scenarios. They can also be used for the calculation of gravitational wave templates for Extreme Mass Ratio Inspirals (EMRI) \cite{BarackCutler07}, for systematically finding homoclinic orbits \cite{LevinPeresGill09}, also in generalized scenarios, for analytic solutions and systematic studies of effective one--body equations of motion \cite{DamourJaranowskiSchaefer00}, for testing numerical codes for binary systems, and finally may also be of use for practical applications like geodesy.  

Analytic solutions of the geodesic equation were only known for the motion in Schwarzschild \cite{Hagihara31}, Reissner--Nordstr\"om, Kerr, and Kerr--Newman space--times \cite{Chandrasekhar83}. Based on new techniques developed in \cite{HackmannLaemmerzahl08} we derived the complete set of analytical solutions of the geodesic equation in Schwarzschild--(anti-)de Sitter space--time and applied it also to the case of spherically symmetric space--times in higher dimensions \cite{Hackmannetal08}. Here we present complete sets of analytic solutions for a wide class of stationary axially symmetric gravitational fields. 

The metric of a general stationary axially symmetric black hole gravitational field in four dimensions 
has the form
\begin{eqnarray} \label{basicmetric}
\rm{d}s^2 & = & \frac{\Delta_r}{p^2} \left(\rm{d}t - A \rm{d}\varphi\right)^2 - \frac{p^2}{\Delta_r} \rm{d}r^2 \nonumber\\
& & - \frac{\Delta_\vartheta}{p^2} \sin^2\vartheta (a \rm{d}t - B \rm{d}\varphi)^2 - \frac{p^2}{\Delta_\vartheta} \rm{d}\vartheta^2 \, , 
\end{eqnarray}
where the functions $p^2$, $\Delta_r$, $\Delta_\vartheta$, $A$, and $B$ will be specified later. We derive all analytic solutions of the geodesic equation in various space--times of this type, namely Taub--NUT--(anti-)de Sitter (TNdS), Kerr--(anti-)de Sitter (KdS) and even in  Pleba\'nski--Demia\'nski (PD) black hole space--times without acceleration. The set of PD black-hole solutions exhausts all electrovac type D spacetimes. Since the Hamilton--Jacobi equation is separable only in the subset of space-times without acceleration \cite{DemianskiFrancaviglia81}, we conclude that we can analytically solve the geodesic equation in all electrovac type-D space-times without acceleration.

In all stationary axisymmetric space--times we have a conserved energy and angular momentum 
\begin{align}
E & = \frac{\Delta_r}{p^2} (\dot t - A \dot\varphi) - a \frac{\Delta_\vartheta}{p^2} \sin^2\vartheta (a \dot t - B \dot\varphi) \\
L & = - A \frac{\Delta_r}{p^2} (\dot t - A \dot\varphi) + B \frac{\Delta_\vartheta}{p^2} \sin^2\vartheta (a \dot t - B \dot\varphi) \, ,
\end{align}
where the dot denotes the derivative with respect to $s$. 

\paragraph{Orbits in Taub--NUT--(anti-)de Sitter space--times}\label{sec:nds} 

The stationary axisymmetric TNdS space--time is given by \cite{Misner63,GriffithsPodolsky06}
\begin{align}
p^2 &: = r^2 + n^2  \,, \;\; A := 2 n \cos\vartheta 
\,, \;\; B := p^2 \,, \;\; \Delta_\vartheta := 1\\
\Delta_r & := r^2 - 2 M r - \frac{\Lambda}{3}  \left(r^2 + 3 n^2\right)^2 - n^2 \left(1 - 4 \Lambda n^2\right),
\end{align}
where $\Lambda$ is the cosmological constant and $M$ and $n$ are the mass and NUT charge of the gravitating body.


With dimensionless quantities $\tilde r = r/r_{\rm S}$ ($r_{\rm S} = 2 M$), $\tilde n = n/r_{\rm S}$, $\tilde L = L/r_{\rm S}$, and the substitution $\xi = \cos\vartheta$ the Hamilton--Jacobi equation separates and reduces to 
\begin{equation}
\frac{\rm{d}\tilde r}{\rm{d}\tau} = \sqrt{R}\, , \quad \frac{\rm{d}\xi}{\rm{d}\tau} = \sqrt{\Theta_\xi}\, , \quad \frac{\rm{d}\varphi}{\rm{d}\tau} = \frac{\tilde L - 2 \tilde n E \xi}{1 - \xi^2} \label{r-eq}
\end{equation}
where
\begin{eqnarray}
R & = & (\tilde r^2+ \tilde n^2)^2 E^2 - \tilde\Delta_r (\delta \tilde r^2 + \tilde L^2 + k) \\
\Theta_\xi & = & \alpha \xi^2 + \beta \xi + \gamma \, .
\end{eqnarray} 
Here, $k$ is the separation constant (Carter constant), $\alpha = - (k - \delta \tilde n^2 + 4 E^2 \tilde n^2 + \tilde L^2)$, $\beta = 4 E \tilde n \tilde L$, $\gamma = k - \delta \tilde n^2$, and $\tilde\Delta_r = \Delta_r r_{\rm S}^{-2}$. Here and throughout the paper we use the Mino--time $\tau$ defined through $p^2 d\tau = r_{\rm S} ds$ \cite{Mino03}. $\delta=1$ for time--like and $\delta = 0$ for null geodesics. 

For given $\Lambda$ and $n$, the orbits are characterized by $E$, $L$, and $k$. The range of these parameters is restricted by the condition that $R$ and $\Theta_\xi$ have to be positive or, equivalently, possess a certain number of real zeros.

\begin{figure}[t]
\subfigure[][escape orbit crossing $r = 0$; $E^2=0.99515$, $k =-0.5$, $\tilde L = 2.5$.]{\label{orb-k-05la25}\includegraphics[width=0.23\textwidth]{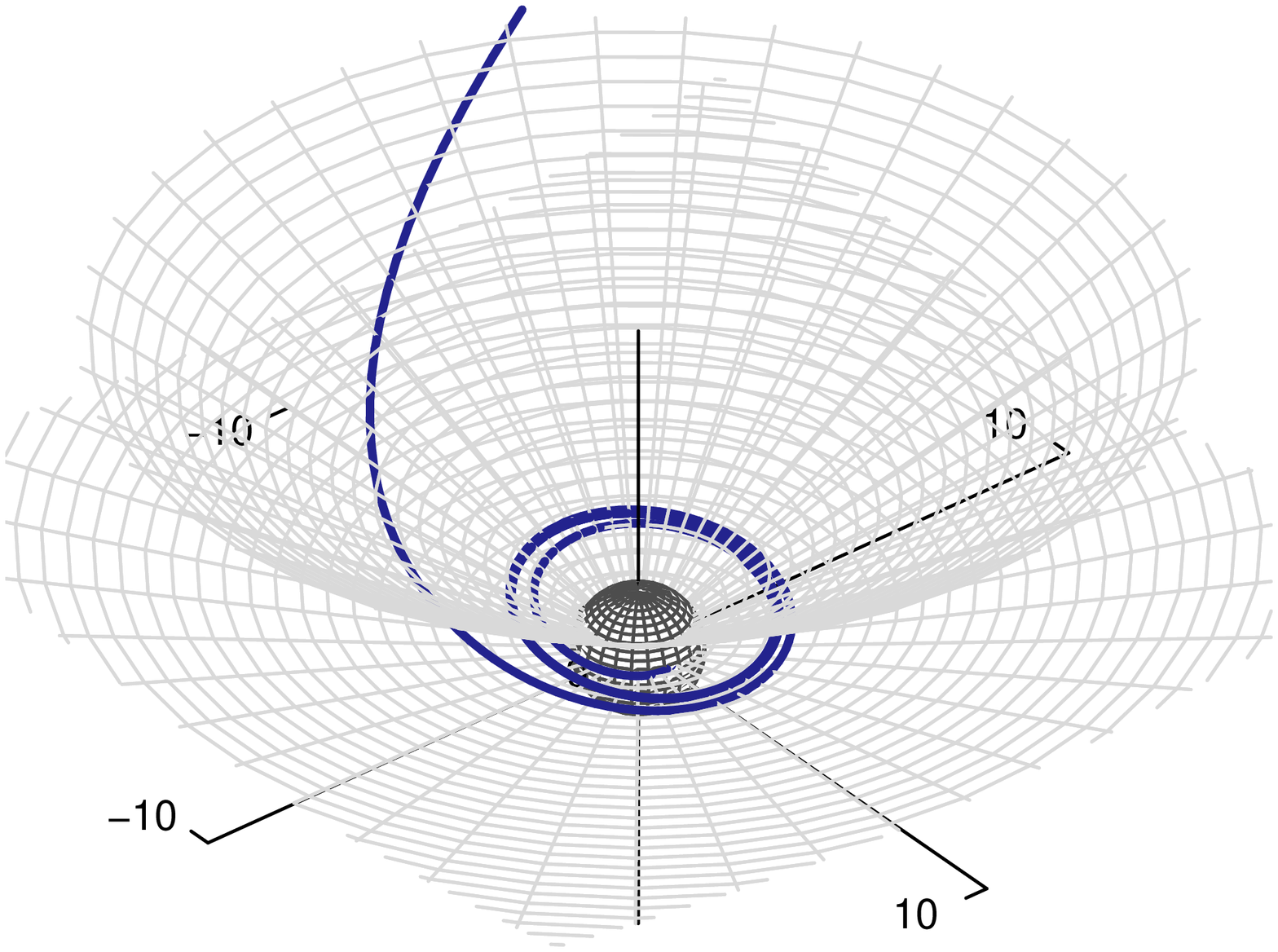}} \,\,
\subfigure[][bound orbit; $E^2 = 0.94038$; $ k =1.0$, $\tilde L =2.0$]{\label{orb-k1la2}\includegraphics[width=0.23\textwidth]{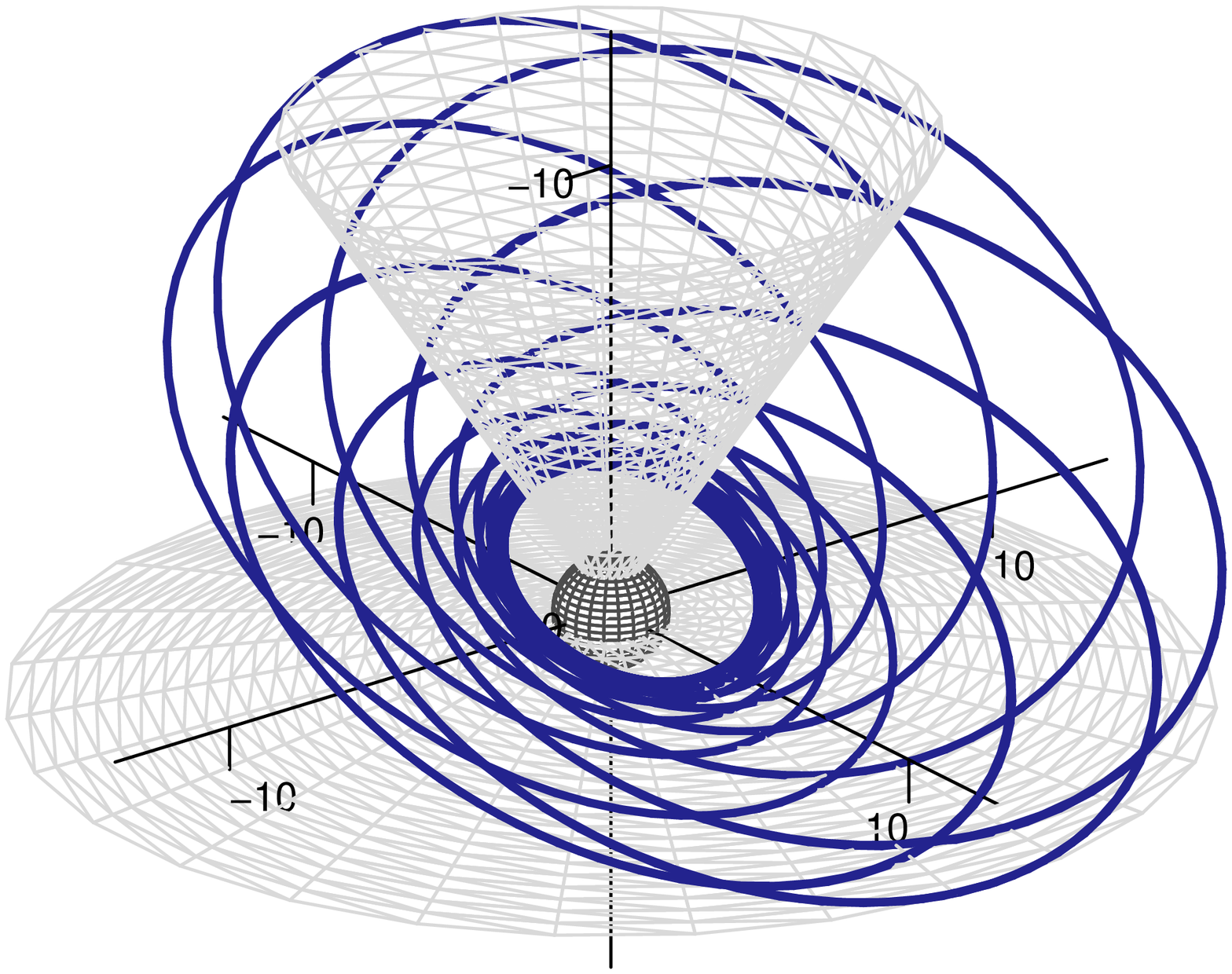}}
\vspace*{-0.2cm}
\caption{TNdS: Orbits for $n=0.5$, $\tilde\Lambda = 2.9\cdot10^{-5}$.} \label{orb-n05La-5}
\end{figure}

$\Theta_\xi$ has two real zeros $\xi_{1,2} \in (-1,1)$ and, thus, can be positive only if
\begin{equation}
\begin{aligned}
c_1 & = k - \delta \tilde n^2  + 4 E^2 \tilde n^2 & \geq 0 \\
c_2 & = k - \delta \tilde n^2 + \tilde L^2  &  \geq 0  
\end{aligned} \label{theta-cond-lamu} 
\end{equation}
which constrains the allowed values for $E$ and $\tilde L$. Then $\xi$ is restricted to the interval $[\xi_1, \xi_2]$ which for $n \neq 0$ is not symmetric with respect to $\xi = 0$. Thus, the motion of a test particle in TNdS space--time is bounded by two cones with in general different opening angles~\cite{LyZon98}. If $k - \delta \tilde n^2 > 0$ then $\vartheta \in (0, \pi)$ (see fig.\ref{orb-k1la2}), but for $k - \delta \tilde n^2 \leq 0$ the motion takes place in one half--space ($\vartheta \in (0, \frac{\pi}{2})$ or $\vartheta \in (\frac{\pi}{2}, \pi)$) only (see Fig.~\ref{orb-k-05la25}). If $n = 0$ or $L = 0$ the opening angles of the two cones are equal and the motion is symmetric with respect to the equatorial plane. It also can be shown along the lines of \cite{LyZon98} that all test particles move on a cone which is tilted with respect to the symmetry axis and which touches the two bounding cones related to $\xi_1$ and $\xi_2$, see fig.~\ref{orb-n05La-5}. 

The function $R$ is a polynomial of $6^{\rm th}$ order in $r$ which for positive $\Lambda$ possesses at most 4 positive real zeros. The $r$ coordinate may take negative values. 
 
\begin{figure}[t!]
\subfigure[][$k =-0.5$]{\label{lamu-n05k-05La-5}\includegraphics[width=0.21\textwidth]{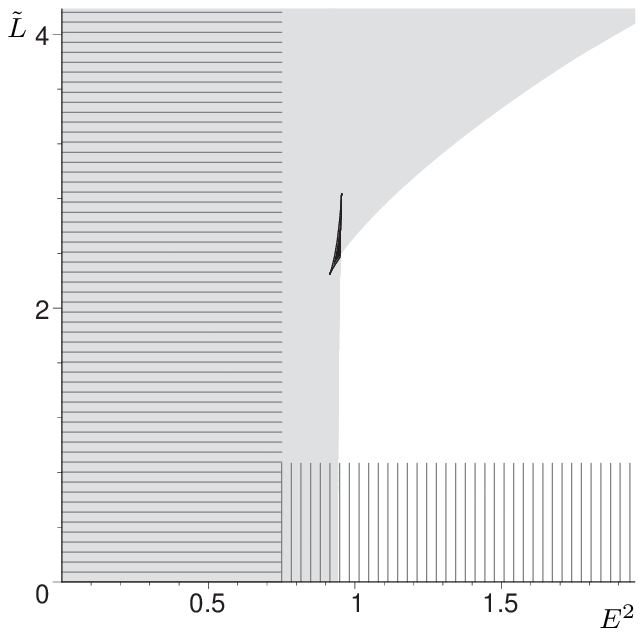}} \,\,
\subfigure[][$k =1.0$]{\label{lamu-n05k1La-5}\includegraphics[width=0.21\textwidth]{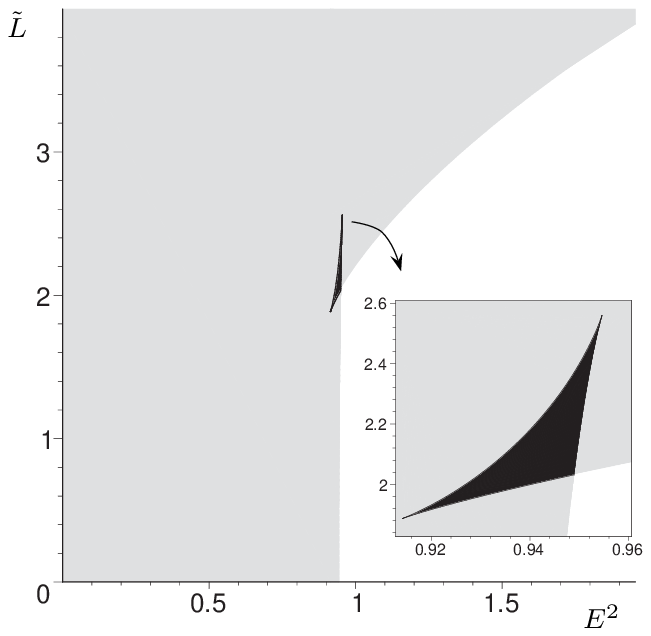}}
\caption{TNdS: Parameter diagrams for $\tilde n = 0.5$, $\tilde\Lambda = 2.9\cdot10^{-5}$. The gray scales denote the number of positive real zeros of the polynomial $R$: black = 4, gray = 2, white = 0. Parameters in the dashed region are forbidden due to \eqref{theta-cond-lamu}.
}\label{lamu-n05La-5}
\end{figure}

Two examples of $(E, \tilde L)$--diagrams for $n=0.5$ and different $k$ are shown in fig.~\ref{lamu-n05La-5} where the dashed region denote forbidden values for $E$ and $\tilde L$ due to~\eqref{theta-cond-lamu}. Black indicates an escape orbit, a bound orbit (\textit{e.g.} fig.~\ref{orb-n05La-5}\subref{orb-k1la2}) and a bound orbit crossing $r = 0$, gray one escape and a bound orbit crossing periodically $r = 0$, and white an escape orbit crossing $r = 0$ (fig.~\ref{orb-n05La-5}\subref{orb-k-05la25}). 

With~\eqref{theta-cond-lamu} we have $\alpha < 0$ and $D := \beta^2 - 4 \alpha \gamma > 0$ and the solution for $\vartheta$ is
\begin{equation}
\vartheta(\tau) = \arccos \Bigl(\frac{1}{2\alpha}\left(\sqrt{D}\sin\left(\tau ^\vartheta_0 - \sqrt{- \alpha} \tau\right) - \beta\right)\Bigr) \ ,
\end{equation}
where $ \tau^\vartheta_0 = \sqrt{- \alpha} \tau_0 + \arcsin\frac{2 \alpha \xi_0 + \beta}{\sqrt{D}}$. 
A substitution $\tilde r = \pm \frac{1}{x} + \tilde r_R$, where $\tilde r_R$ is a zero of $R$, reduces the equation for $\tilde r$ in \eqref{r-eq} to the hyperelliptic differential of first kind $d\tau = \frac{x dx}{\sqrt{P_5(x)}}$ for a polynomial $P_5$ of degree 5. This can be solved by \cite{HackmannLaemmerzahl08}
\begin{equation}
\tilde r(\tau) = \mp \frac{\sigma_2}{\sigma_1}\begin{pmatrix} f(\tau - \tau^\prime_0) \\ \tau - \tau^\prime_0 \end{pmatrix} + \tilde r_R \ , \label{r-nds}
\end{equation}
where $\sigma_i$ are the derivatives of the Kleinian $\sigma$ function, $f$ describes the $\theta$--divisor, \textit{i.e.} $\sigma((f(z),z)^t) = 0$, and $\tau^\prime_0 = \tau_0 + \int^\infty_{x_0} \frac{x dx}{\sqrt{P_5(x)}}$. The solution of the $\varphi$ equation is
\begin{eqnarray}
\varphi(\tau) & = & \varphi_0 + \frac{1}{2} \Biggl(A_+ \arctan\frac{1- u B_-}{\sqrt{1-u^2}\sqrt{B_-^2 - 1}} \nonumber \\
&& - A_- \arctan\frac{1-u B_+}{\sqrt{1 - u^2} \sqrt{B_+^2 - 1}}\Biggr) \Biggl|^{\xi(\tau)}_{\xi_0} \label{sol2phi} \ ,
\end{eqnarray}
where $u = \frac{2 \alpha \xi + \beta}{\sqrt{D}}$, $A_\pm = \frac{\tilde L \pm 2 E \tilde n}{|\tilde L \pm 2 E \tilde n|}$ and $B_\pm = \frac{\beta \pm 2 \alpha}{\sqrt{D}}$. 

For a detailed discussion see~\cite{nds09}.

\paragraph{Orbits in Kerr--(anti-)de Sitter space--times}\label{sec:kerrds}

The stationary axially symmetric KdS metric written in Boyer--Lindquist coordinates $t \rightarrow t \chi^{-1}$, $\varphi \rightarrow \varphi \chi^{-1}$, where $\chi = 1 + \frac{1}{3} a^2 \Lambda$, emerges from~\eqref{basicmetric} for \cite{Carter68,GriffithsPodolsky06}
\begin{align}
p^2 & := r^2 + a^2 \cos^2\vartheta \,,\quad \Delta_\vartheta := 1 + \tfrac{1}{3} \Lambda a^2 \cos^2\vartheta \,,\\
\Delta_r & := \left(1 - \tfrac{1}{3} \Lambda r^2\right) (r^2+a^2) -2Mr \,,\\
A & := a \sin^2\vartheta  \,, \quad B := r^2+a^2 \,.
\end{align}

The Hamilton--Jacobi equation again separates and gives with $\tilde a = a/r_{\rm S}$ and $\mathcal{D} = \tilde L/E$, $\kappa = k/E^2$, $\delta_2 = \delta/E^2$ and the substitution $\nu = \cos^2\vartheta$ (we exclude $\nu=1$ corresponding to the coordinate singularities $\vartheta =0,\pi$)
\begin{align}\label{dotr_theta_sn}
\frac{\rm{d}\tilde r}{\rm{d}\tau} & = E  \sqrt{R} \, , \quad \frac{\rm{d}\nu}{\rm{d}\tau} = E  \sqrt{\nu \Theta_\nu} 
\\
\label{dotphi_sn}
\frac{\rm{d} \varphi}{\rm{d} \tau} & = \tilde{\chi}^2 E \left[\frac{\tilde a \left(\tilde r^2 + \tilde a^2 - \tilde a \mathcal{D}\right)}{\tilde\Delta_r} - \frac{\tilde a \sin^2\vartheta - \mathcal{D}}{\tilde\Delta_\vartheta \sin^2\vartheta}\right] \, ,
\end{align} 
where
\begin{align}
R & := \chi^2 (\tilde r^2 + \tilde a^2 - \tilde a \mathcal{D})^2 - \tilde\Delta_r (\delta_2 \tilde r^2  + \kappa) \label{defR} \\
\Theta_\nu & := (1 - \nu) \Delta_\vartheta (\kappa - \delta_2 \tilde a^2 \nu) - \chi^2 \left(\tilde a (1 - \nu) - \mathcal{D}\right)^2 \label{defTheta}
\end{align}

Solutions of \eqref{dotr_theta_sn} require $\Theta_\nu(\nu) \geq 0$ and $R(\tilde r) \geq 0$. Therefore, the number of real zeros of $\Theta_\nu$ (in $[0, 1)$) and $R$ depending on the parameters of the black hole and the particle gives all possible types of orbits. 

The number of zeros of $\Theta_\nu$ in $[0,1)$ changes if (i) $0$ or $1$ is a zero or (ii) two zeros coincide. We have case (i) if $\tilde L  = \tilde a E \pm \frac{\sqrt{k}}{\chi}$ or $\tilde L = 0$, respectively. Considering case (ii) we can solve $\Theta_\nu = (\nu - u)^2 (a_1 \nu + a_0)$ with some constants $a_i$ for $E^2$ and $\tilde L$, where only $u \in [0,1)$ is of interest. In the same way we can analyze the changes in the number of zeros of $R$ with the ansatz $R = (\tilde r - u)^2 (\sum_{i=0}^4 a_i \tilde r^i)$.   

\begin{figure}
\includegraphics[width=0.23\textwidth,trim=9 9 9 9, clip]{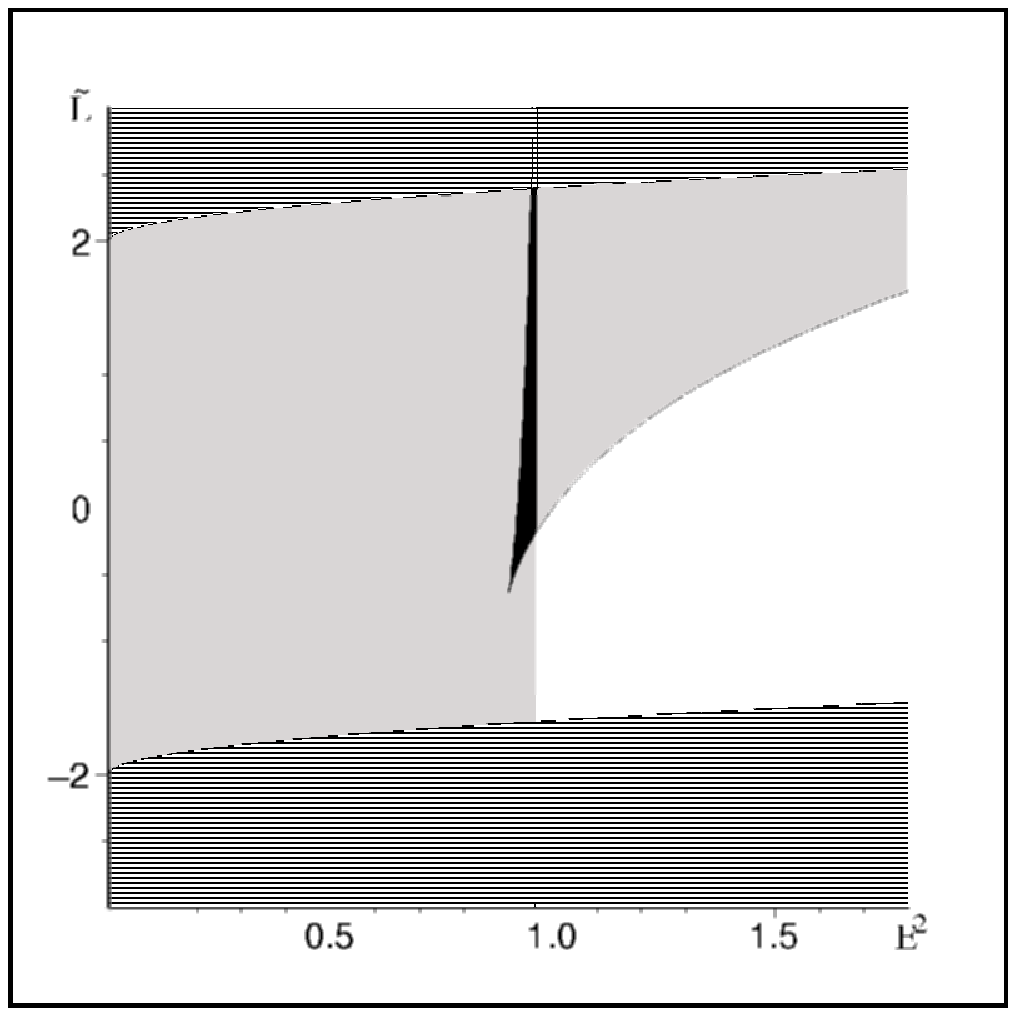}
\caption{Different types of geodesic motion. Here $r_{\rm S}=2$, $\tilde a =0.4$, $k=4$, and $\tilde\Lambda  =10^{-5}$.  The gray scales denote the number of positive real zeros of the polynomial $R$: black = 5, gray = 3, white = 1. Parameters in the dashed region are forbidden due to $\Theta_\nu \geq 0$.}
\label{Fig:KdStypes}
\end{figure}

A typical result of such an analysis for timelike geodesics is shown in fig.~\ref{Fig:KdStypes}. The dashed region denotes forbidden parameter values. The white area corresponds to an escape orbit, the gray area to an escape orbit but also a bound orbit, and the black area to an escape and two different bound orbits. 

The differential equation for $\vartheta$ is of elliptic type and first kind and is solved by
\begin{equation}
\vartheta(\tau) = \arccos \pm \sqrt{\frac{d_1}{4 \wp(2 E \tau - \check{\tau}_0 ;g_2,g_3) - d_2}} \, , 
\end{equation}
where $g_2 = \frac{3}{4} d_2^2 - d_3 d_1$, $g_3 = \frac{1}{4} d_1 d_2 d_3 - \frac{1}{16} \delta_2 \tilde a^4 d_1^2 - \frac{1}{8} d_2^3$, $\check{\tau}_0 = 2 E \tau_0 + \int_{y_0}^\infty \frac{dy'}{\sqrt{4 {y'}^3 - g_2 y'- g_3}}$, and $d_1 = \kappa - \chi^2 (\tilde a - \mathcal{D})^2$, $3 d_2 = \tilde a^2 (\kappa \tilde\Lambda  -\delta_2) - \kappa + 2 \chi^2 \tilde a (\tilde a - \mathcal{D})$, $4 d_3 = \tilde a^2 (\delta_2 (1 - \Lambda a^2) - \chi^2 - \tilde\Lambda \kappa)$. Contrary to the TNdS--case, this motion is symmetric with respect to the equatorial plane. 

The differential equation for $\tilde r$ is of hyperelliptic type and first kind. It can be cast in standard form $x \frac{dx}{d\tau}= E \sqrt{a_5 R_x}$ by a substitution $\tilde r=\pm 1/x+\tilde r_R$ where
\begin{equation}
R_x = \sum_{i=0}^5 \frac{a_i}{a_5} x^i \, , \quad a_i = \frac{(\pm 1)^i}{(6-i)!} \frac{d^{(6-i)} R}{d\tilde r^{(6-i)}} (\tilde r_R) \, ,
\end{equation}
and $\tilde r_R$ is a zero of $R$ and $a_5>0$. With $\hat{\tau}_0 = E \sqrt{a_5} \tau _0 + \int_{x_0}^\infty \frac{x dx}{\sqrt{\tilde R_x}}$ the solution is given by \cite{HackmannLaemmerzahl08}
\begin{equation}
\tilde r(\tau) = \mp \frac{\sigma_2}{\sigma_1}\left(\begin{matrix} f(E \sqrt{a_5} \tau - \hat{\tau}_0) \\ E \sqrt{a_5} \tau - \hat{\tau}_0  \end{matrix}\right) + \tilde r_R\, . \label{r-kds}
\end{equation}

\begin{figure}[t]
\subfigure[][escape orbit, $E^2 = 0.98$, $\tilde L = 1$.]{\label{orb-k-05la25kds}
\includegraphics[width=0.23\textwidth]{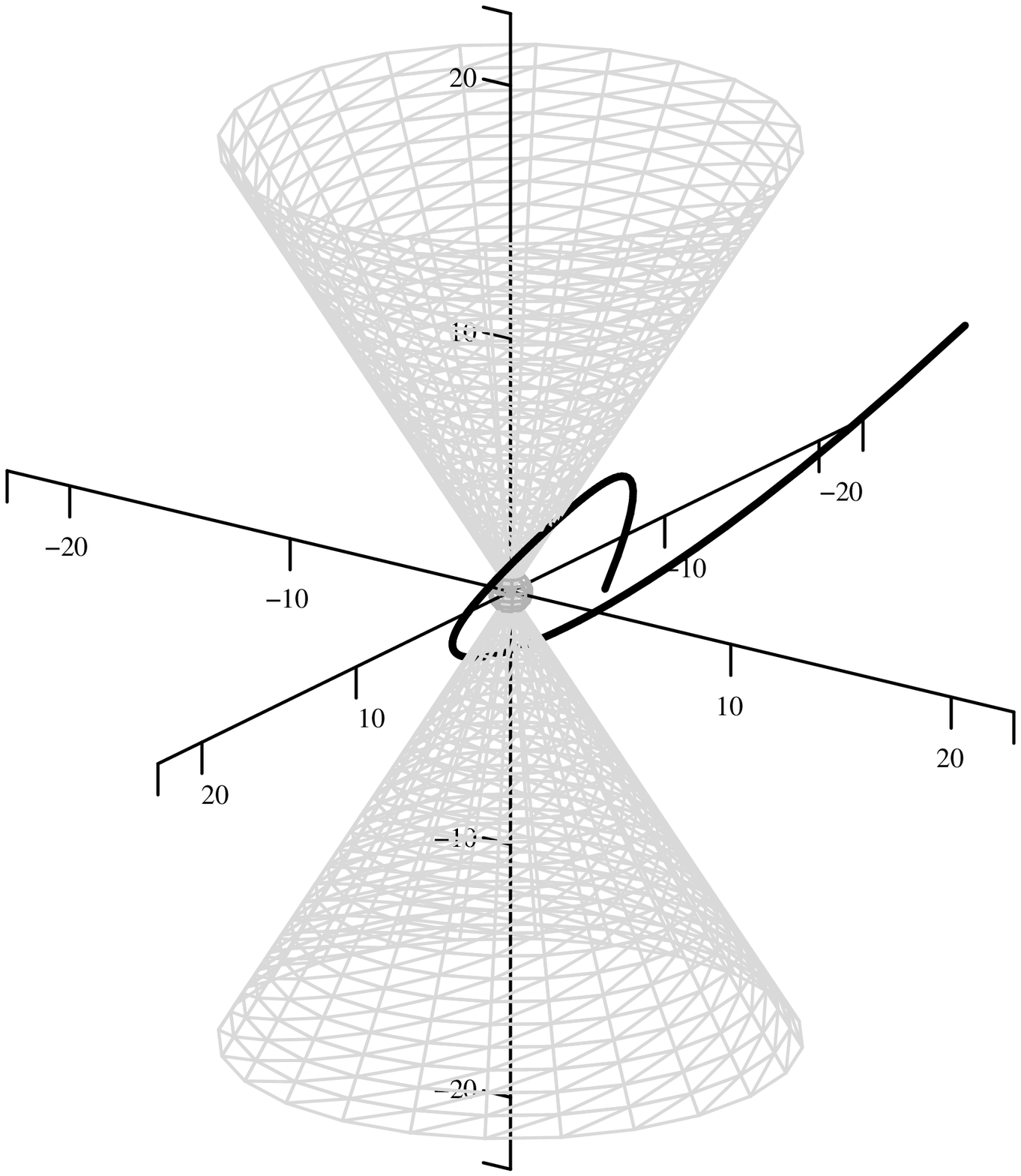}} 
\subfigure[][bound orbit, $E^2 = 0.94$, $\tilde L = 0.6$.]{\label{orb-k1la2kds}\includegraphics[width=0.23\textwidth]{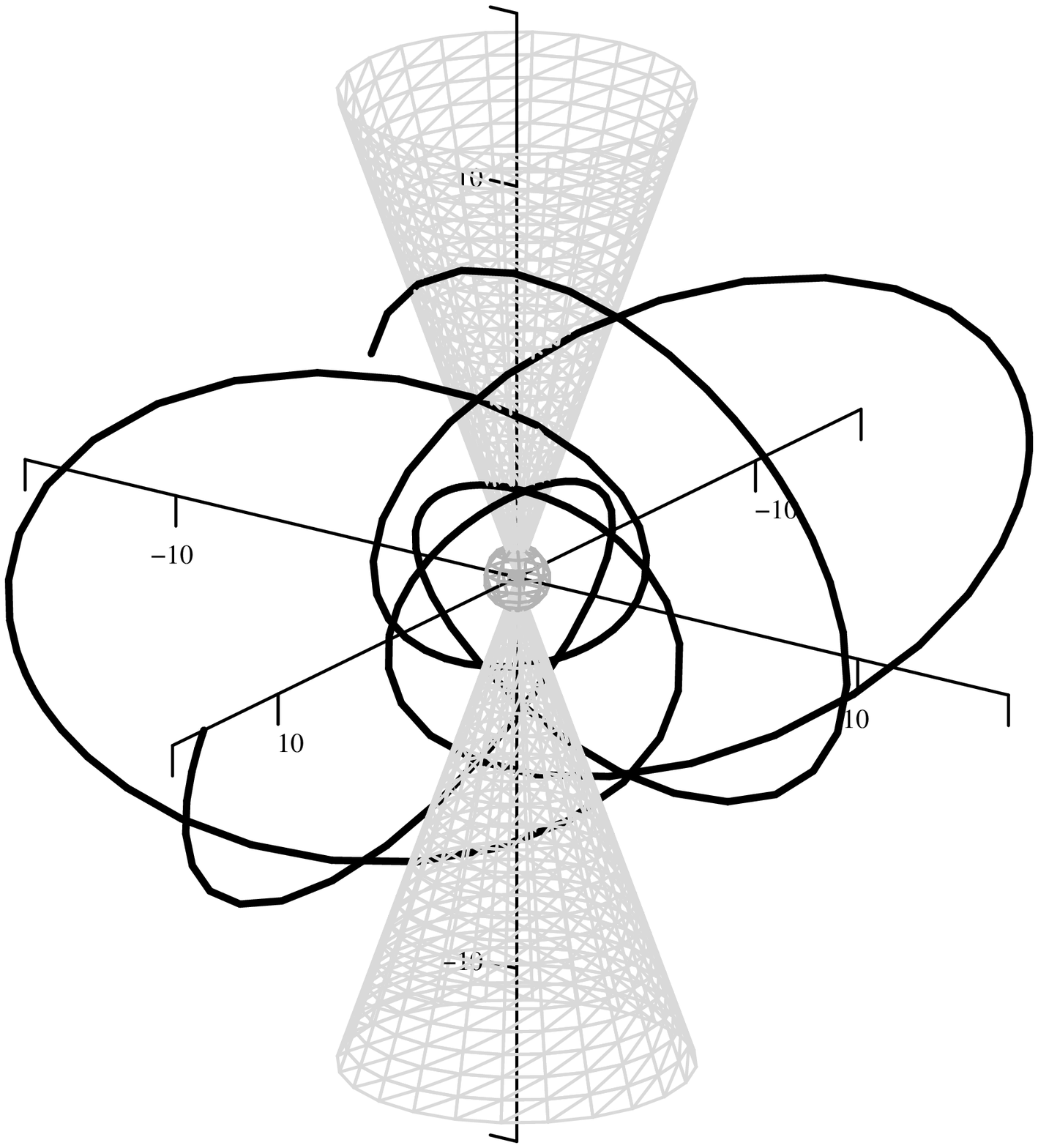}}
\vspace*{-0.2cm}
\caption{KdS: Orbits for $\tilde a = 0.4$, $\tilde\Lambda = 10^{-5}$, and $k = 3$.} \label{orb-kds}
\end{figure}

The $\varphi$--motion depends on both $r$ and $\vartheta$
\begin{equation}
\chi^{-2}(\varphi - \varphi_0) = I_r - I_\vartheta \, ,
\end{equation}
where we have to solve an elliptic and hyperelliptic integral of third kind \cite{nds09}
\begin{align}
&\frac{2 c_1}{|c_1|} I_{\vartheta} = \left(\sum_{i=1}^4 \frac{1}{\wp'(v_i)} \left(\zeta(v_i) (v-v_0) + \log \frac{\sigma(v-v_i)}{\sigma(v_0-v_i)}\right)\right) \nonumber \\
& \quad \cdot \left((\delta_{i1} + \delta_{i2}) \tilde a^3 \tilde\Lambda (\chi - a \Lambda \mathcal{D}) + (\delta_{i3} + \delta_{i4}) \mathcal{D} \right) \frac{(- d_1)}{4 \chi} \nonumber \\
& \quad + (\tilde a - \mathcal{D}) (v-v_0) \, ,
\end{align}
where $v(\tau) = 2 E \tau - \check{\tau}_0$, $v_0 = v(\tau_0)$, $\wp(v_1) = \frac{1}{4} (a^2 \Lambda d_1 - d_2) = \wp(v_2)$, $\wp(v_3)= \frac{1}{4} (d_1 + d_2) = \wp(v_4)$, 
and
\begin{multline}
- \frac{\sqrt{a_5} |x_0|}{\tilde a x_0} I_{\tilde r} = C_1(w-w_0) + C_0(f(w) + f(-w_0)) + \sum_{i=1}^4 \\ 
\frac{C_{2,i}}{\sqrt{R_{u_i}}} \left[ S_i(w)  - S_i(w_0) - \begin{pmatrix} f(w) \\ w \end{pmatrix}^t \left(\int_{p_i^-}^{p_i^+} d\vec r\right)\right] ,
\end{multline}
where $w(\tau) = E \sqrt{a_5} \tau - \hat{\tau}_0 $, $w_0 = w(\tau_0)$, and $S_i(w)= \ \log\frac{\sigma((f(w),w)^t- 2 \int_\infty^{p^{+}_i}d\vec z)}{\sigma((f(w),w)^t - 2 \int_\infty^{p^{-}_i}d\vec z)}$. Here $d\vec z = \left( \frac{dx}{\sqrt{R_x}}, \frac{x dx}{\sqrt{R_x}} \right)^t$ and $d \vec r = (dr_1,dr_2)^t$ with $dr_i = \sum_{k=i}^{5-i} (k+1-i) \frac{a_{k+1+i}}{a_5} \frac{x^k dx}{4 \sqrt{R_x}}$ (see~\cite{BuchstaberEnolskiiLeykin97}). The constants $C_j$ are given by a partial fraction decomposition. The points $p^{\pm}_i = (x_i, \pm \sqrt{R_{x_i}})$ are the zeros of $\Delta_{\tilde r(x)}$ on the Riemann surface of $y^2=R_x$. Two orbits are shown in fig.~\ref{orb-kds}, for a detailed 
discussion see~\cite{nds09}. 

\paragraph{Orbits in Pleba\'nski--Demia\'nski space--times}\label{sec:PD}

The PD space--times without acceleration are described by~\eqref{basicmetric}
with~\cite{GriffithsPodolsky06}
\begin{align}
p^2 & = r^2 + \left(n - a \cos\vartheta \right)^2 \,,\\
\Delta_\vartheta & = 1 + \tfrac{1}{3} a^2 \Lambda \cos^2\vartheta - \tfrac{4}{3} \Lambda a n \cos\vartheta \\
\Delta_r & = r^2 - 2 M r - n^2 + a^2 + Q^2_{\rm e} + Q_{\rm m}^2 \nonumber \\
& \quad - \tfrac{1}{3} \Lambda \left(r^4 + (6 n^2 + a^2) r^2 + 3 (a^2 - n^2) n^2\right)  \\
A & = a \sin^2\vartheta  + 2 n \cos\vartheta  \,, \quad B = r^2 + a^2 + n^2 \,,
\end{align}
where $Q_{\rm e}$ and $Q_{\rm m}$ are electric and magnetic charges of a gravitating source. Also in this case the Hamilton--Jacobi equation separates (for simplicity, we choose neutral test particles; charged particles do not change the structure of the equations) and yields differential equations for $\tilde r$, $\xi = \cos\vartheta$, and $\varphi$ 
\begin{equation}\label{PlebDemeom}
\frac{\rm{d}\tilde r}{\rm{d}\tau} = \sqrt{R} \, , \quad \frac{\rm{d}\xi}{\rm{d}\tau} = \sqrt{\Theta_\xi} \, , \quad \frac{\rm{d}\varphi}{\rm{d}\tau} = \frac{\tilde a {X}}{\tilde\Delta_r} + \frac{\tilde L - \tilde{A} E}{\tilde\Delta_\vartheta \sin^2\vartheta} 
\end{equation}
with
\begin{align}
R & = {X}^2 - \tilde\Delta_r (\delta \tilde r^2 + k)  \\
\Theta_\xi & = \Delta_\vartheta (1 - \xi^2) \left(k - \delta (\tilde n - \tilde a \xi)^2 \right) - (\tilde L - \tilde{A} E)^2  \, , 
\end{align}
where ${X}= (\tilde r^2 + \tilde a^2 + \tilde n^2) E - \tilde a \tilde L$ and ${\tilde{A}} = A r_{\rm S}^{-1}$. 

A standard substitution $\xi=\pm \frac{1}{y} + \xi_\Theta$, where $\xi_\Theta$ is a zero of $\Theta_\xi$, brings the equation for $\xi$ to a holomorphic hyperelliptic differential of the first kind $\rm{d} \tau = \frac{y dy}{\sqrt{P_5(y)}}$. The solution of the equation for $\vartheta$ in \eqref{PlebDemeom} is
\begin{equation} \label{theta-sol-PD}
\vartheta(\tau) = \arccos\Bigg(\mp \frac{\sigma_2}{\sigma_1}\begin{pmatrix} f(\tau - \tau ^{\vartheta}_0) \\ \tau - \tau^{\vartheta}_0 \end{pmatrix} + \xi_\Theta\Bigg)  \,.
\end{equation}
As in \eqref{r-nds} we obtain for the $\tilde r$ motion the solution 
\begin{equation} \label{r-sol-PD}
\tilde r(\tau) = \mp \frac{\sigma_2}{\sigma_1}\begin{pmatrix} f(\tau - \tau^{\tilde r}_0) \\ \tau - \tau^{\tilde r}_0 \end{pmatrix} + \tilde r_R  \,.
\end{equation}
The function $f$ describing the $\theta$--divisor depends on the definition of the $\sigma$--function and, therefore, on the polynomials $P_5(y)$ for the $\vartheta$ motion and on $P_5(x)$ for the $\tilde r$ motion.

$\varphi$ depends both on $\vartheta$ and $\tilde r$ giving a hyperelliptic integral of third kind which can be solved completely \cite{nds09}. 
 
Therefore we succeeded in obtaining the complete analytic solution of the geodesic equation in all PD black hole space--times without acceleration. The orbits now depend on the three particle parameters $E$, $L$, and $k$ as well as the six parameters characterizing this class of space--times. It has been shown that the PD black hole solutions exhaust all electrovac type D solutions. Since the condition of a space--time of being of electrovac type D without acceleration ensures separability of the Hamilton--Jacobi equation \cite{DemianskiFrancaviglia81} we, thus, arrived at the conclusion that we now can explicitly give all analytic solutions of geodesic equations in all electrovac type D space--times without acceleration. 

In all cases discussed here the motion can be extended through $r = 0$ to negative $r$ \cite{nds09}.

For bound orbits, the $r$ and $\vartheta$ motions are related to characteristic periods given by 
\begin{equation}
\omega_r = 2 \int_{r_{\rm min}}^{r_{\rm max}} \frac{dr}{\sqrt{R(r)}} \, , \quad \omega_\vartheta = 2 \int_{\xi_{\rm min}}^{\xi_{\rm max}} \frac{d\xi}{\sqrt{\Theta_\xi(\xi)}} \, ,
\end{equation}
which are related to the zeros of the underlying polynomials. From these periods one can derive the perihelion shift and the Lense--Thirring effect, and also perform further orbital frequency analysis \cite{DrascoHughes04}. For escape orbits, the two deflection angles are given by $\displaystyle\Delta\vartheta = \lim_{s \rightarrow \infty} \left(\vartheta(s) - \vartheta(-s)\right)$ and  $\displaystyle\Delta\varphi = \lim_{s \rightarrow \infty} \left(\varphi(s) - \varphi(-s)\right)$. These observable periods and deflection angles can be calculated analytically. For detailed calculations and discussions see \cite{nds09}. 

\begin{acknowledgements}
We are grateful to W. Fischer and P. Richter for helpful discussions. E.H. thanks the German Research Foundation DFG and V.K. the German Academic Exchange Service DAAD for financial support.
\end{acknowledgements}

\end{document}